\documentclass{cernrep}
\usepackage{multirow}
\usepackage{textcomp}
\usepackage{hhline}
\usepackage{tabularx}
\usepackage[flushleft]{threeparttable}
\usepackage[table]{xcolor}

\usepackage{xcolor}
\definecolor{light-gray}{gray}{0.8}
\definecolor{apsblue}{rgb}{0.176, 0.152, 0.57}
\usepackage[colorlinks=true, linkcolor=black, citecolor=apsblue, filecolor=black, urlcolor=apsblue]{hyperref}

\usepackage{caption}
\usepackage{subcaption}
\usepackage{comment}
\usepackage{hyperref}
\usepackage{siunitx}
\newcommand{\murm}{\hbox{\textmu}}
\newcommand{\ee}{$\rm{e}^+\rm{e}^-$}
\newcommand{\tabindent}{\hspace*{0.5cm}}
\begin{document}

\thispagestyle{empty}

\begin{centering}
{\Large {\bf HALHF: a hybrid, asymmetric, linear Higgs factory\\ using plasma- and RF-based acceleration}}\\
\end{centering}
\vspace{60pt}
\begin{figure}[h]
	\centering\includegraphics[width=0.2\textwidth]{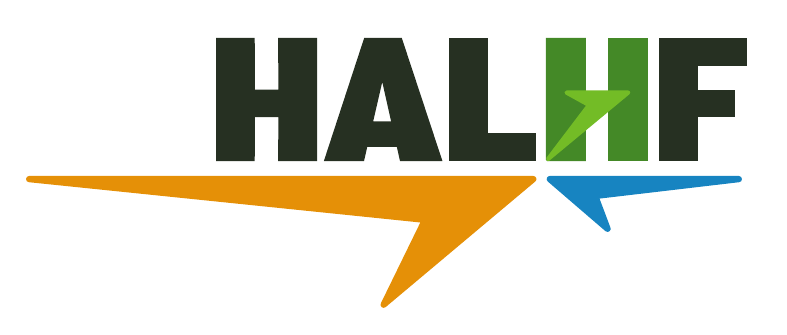}
\end{figure}

\vspace{60pt}
\hspace{-31pt}
\textit{Contact persons:} \\
Brian Foster, University of Oxford/DESY, brian.foster@physics.ox.ac.uk\\
Richard D'Arcy, University of Oxford, richard.darcy@physics.ox.ac.uk\\
Carl A. Lindstr{\o}m, University of Oslo, c.a.lindstrom@fys.uio.no\\

\begin{centering}
\vspace{80pt}
{\large{\bf Abstract}}\\
\vspace{20pt}
\end{centering}
\hspace{-31pt}
HALHF is a hybrid linear collider that uses electron-driven plasma-wakefield acceleration to accelerate electrons to high energy while using radio-frequency cavity technology to accelerate positrons. The most cost-effective solution collides low-energy positrons with high-energy electrons, producing a boost to the final state in the electron direction with $\gamma= 1.67$. The current HALHF baseline design produces a luminosity comparable to that of the baseline ILC but with a greatly reduced construction and carbon footprint and hence much lower cost than the mature linear-collider designs ILC and CLIC. Costs for HALHF are evaluated, together with that for the approximate 15-year R\&D programme necessary to realise HALHF. Time scales and cost for the R\&D are estimated. Upgrade paths for HALHF technology from a 250~GeV Higgs factory, through 380 and 550~GeV, up to 10~TeV are sketched.

\title{HALHF: a hybrid, asymmetric, linear Higgs factory\\ using plasma- and RF-based acceleration}
\author{The HALHF Collaboration$^{\dagger}$}
\maketitle
\thispagestyle{plain}
\setcounter{page}{1}
\vspace{-70pt}
\section{Scientific context}
The previous EPPSU concluded that the next major particle-physics facility should be an $\rm{e}^+\rm{e}^-$ Higgs Factory; it also prioritised R\&D in accelerator science~\cite{CERN_Strategy_2020}. Plasma-based acceleration is an emerging accelerator technology that has shown tremendous progress over the past decade, to the extent that, for the first time, beam-driven plasma-wakefield acceleration (PWFA) approaches some of the parameters required to produce a viable high-energy \ee\ collider for particle physics~\cite{Darcy2022,Lindstrom2021,Lindstrom2024,Pena2024}.

A Higgs factory based on plasma acceleration promises radically to reduce the physical, economic and environmental footprint required to construct a Higgs factory facility. Furthermore, plasma-wakefield technology, whether beam-driven or laser-driven (LWFA) has a plethora of applications across many fields of science. The application to particle physics is by far the most difficult and distant from the current state-of-the-art, so that developments towards a particle-physics collider are likely to have significant spin-offs and a variety of off-ramps to important applications in other fields. 

One crucial and until recently show-stopping impediment to PWFA collider applications is related to the inherent asymmetry of plasmas. These are composed of light, mobile, negatively charged electrons and heavy, immobile, positive ions. While highly performant electron acceleration is now relatively routine, positron acceleration is problematic. HALHF~\cite{HALHF, HALHF_upgrades} circumvents this problem by using RF technology to accelerate positrons, while reserving PWFA for what it does best, accelerating electrons.  

\noindent The advantages of HALHF are:
\begin{itemize}
\item  comparable luminosity and physics sensitivity to e.g. ILC at much smaller capital cost and associated carbon footprint with comparable running costs;  
\item an upgrade path to the TeV energy range and beyond, utilising experience gained at HALHF to increase gradients giving access to \ee\ beyond 1 TeV and $\gamma \gamma$ or $\rm{e}^- \rm{e}^-$ up to 10 TeV;
\item HALHF technology can be used to boost the energy of an already existing linear collider cheaply;
\item the specificity of HALHF’s parameters catalyses R\&D which will enable applications of substantial societal benefits and economic impact well beyond particle physics: e.g. a plasma accelerator with HALHF performance could drive a free-electron laser; using a laser instead of a beam driver, FELs could be viable on sites of medium-sized universities. High-energy electrons from plasma accelerators would also be useful for medical applications and strong-field QED experiments. 
\end{itemize}
HALHF's disadvantage is that it is less technologically mature and more complex than other options and therefore resources for a substantial programme of R\&D are required.

The general scientific case in particle physics for HALHF is similar to that for other linear-collider Higgs factories, and is made in the LCVision submission~\cite{LCVision}. The separation of running into various stages and the flexibility is also identical to that set out in LCVision. Potential differences in the physics performance of HALHF from that of a symmetrical linear-collider Higgs factory, e.g. because of the boosted final state, are commented on in this submission.

\section{Basis and Development of the HALHF proposal}
\label{sec:BandD}
The HALHF concept aims to utilise the orders-of-magnitude higher gradients obtainable in a plasma wakefield compared to RF cavities to propose a disruptive Higgs Factory that is much cheaper in capital cost than any other proposal. While PWFA has been shown to accelerate electrons with gradients up to 100~GV/m~\cite{Corde2016}, and there has been substantial progress in many areas towards the parameters required for particle-physics applications, the major difficulty that has stalled progress in developing a plasma-based linear \ee\ collider has been the asymmetric nature of plasmas. This means that currently no viable technique to accelerate positrons with excellent beam quality and high efficiency, as quantified by the luminosity-per-power ratio, is known~\cite{Cao2024}.

The HALHF concept side-steps this problem by using   RF  structures to accelerate positrons. To minimise cost, the energy is substantially reduced compared to that of the electron, exploiting  the cost-effectiveness of PWFA technology. The optimisation of a collider with such an energy asymmetry favours asymmetry between electron and positron beams in other parameters, e.g.\ bunch currents. Requirements of power efficiency increase the positron bunch current with respect to that of the electrons. 

The original HALHF concept~\cite{HALHF} has been refined~\cite{Foster2025} to incorporate new developments and optimise the physics potential. Firstly, the turn-around loops for the positron bunch and the drive beams in the folded design conceived to minimise the HALHF footprint were too tight to maintain the requisite beam emittances and did not scale favourably to energies beyond a Higgs factory. Secondly,
the dual-purpose linac that accelerated both the colliding positrons and the electron bunch-train required to drive the PWFA arm proved difficult, although not impossible, to design.

A new baseline has now been approved by the HALHF Collaboration, which not only addresses these difficulties but also adds new capabilities, including positron polarization and two interaction points (IP). Positron polarization was previously only an option but the physics gain implied by polarization of both leptons was judged sufficiently significant to outweigh the complications inherent in producing polarized positrons. Although HALHF could be built at many sites, in particular because of its small footprint in comparison to other Higgs factories, the costings and some other aspects given in this document assume a CERN site, \textit{viz.}\ that already investigated for CLIC; this allows more accurate cost estimates than a site-independent design and easier comparison to other projects. 

\section{Current Status of the HALHF proposal}
\label{sec:CS}
\begin{figure}[t]
\vspace{-1cm}
	\centering
    \includegraphics[width=\textwidth]{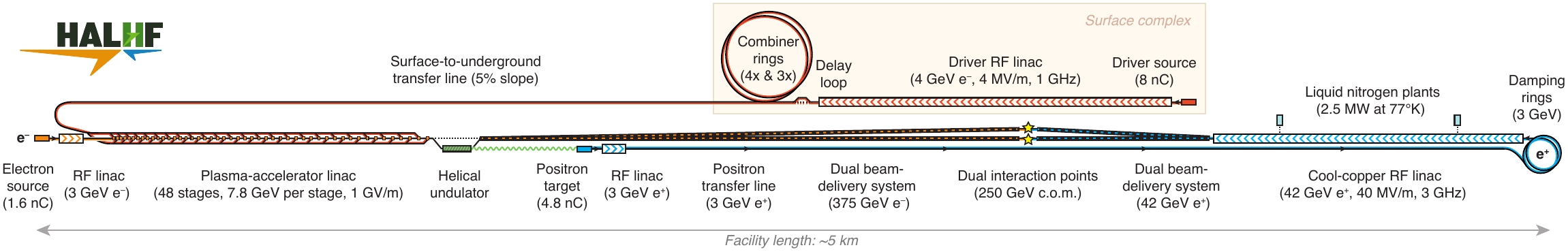}
\caption{Schematic view of the new HALHF baseline at 250~GeV CoM. The red sections relate to electrons, blue to positrons and green to photons. Other components are as labelled on the figure.}
    \label{Fig-1}
\end{figure}

\begin{figure}[t]
	\centering\includegraphics[width=\textwidth]{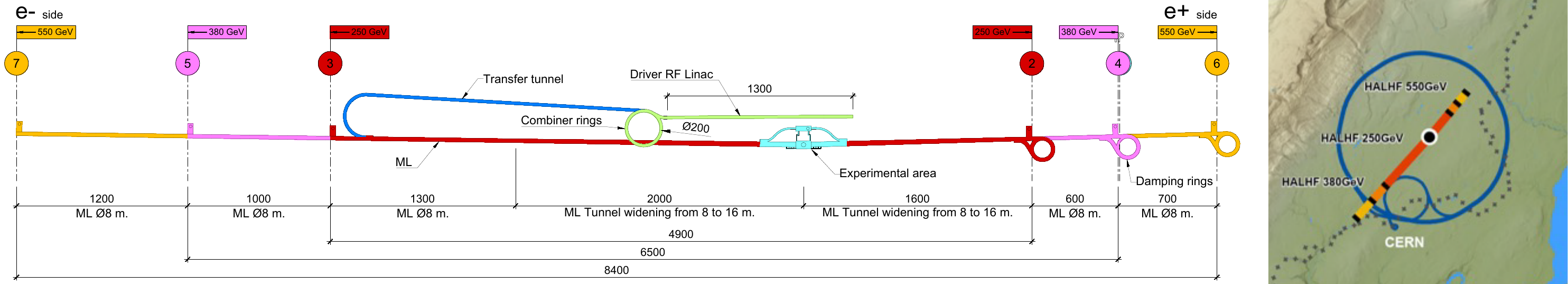}
\caption{Schematic view of HALHF infrastructure (left) and a potential siting at CERN (right).}
    \label{Fig-civilstudy}
\end{figure}

The proposed layout for HALHF is shown in Fig.~\ref{Fig-1}. The electron bunch is produced by a high-quality polarised source and accelerated in the PWFA arm by 48 drive-bunches generated in a facility very similar to that used for CLIC. The colliding bunch produces photons in an undulator that generate polarized positron bunches that are transported to damping rings at the other end of the facility. The damped positrons are accelerated in a Cool Copper linac. Electron and positron bunches at full energy are each directed alternately into two beam-delivery systems (BDS) and collide at separate interaction points. A schematic of how HALHF might appear at a CERN site is shown in Fig.~\ref{Fig-civilstudy}.

The production of positrons is based on that of ILC. The colliding electron beam at full energy is passed through an undulator; the resulting photons impinge on a rotating target. The positrons from the resulting \ee\ pairs are concentrated and collected. With appropriate modifications to the undulator parameters, it is expected that a high degree of polarization can be achieved~\cite{Foster2025}.

Separating the drive-beam and positron linacs provides wider scope for optimisation of the overall HALHF facility. In particular, allowing different energy for the drive-beams and the colliding positron beam is important. The enlarged parameter space was optimised using a multi-dimensional Bayesian optimisation programme. Input to this was a detailed cost model, details of which are given in the Backup; some parts of the model remain preliminary but should not affect the optimisation process. The metric used in the optimisation consisted of the sum of separate costs for construction, overheads, running costs, maintenance over the duration of the full programme and a ``carbon tax'' to give weight to minimising the carbon footprint. The full programme cost was optimised on the assumption of delivering a total integrated luminosity of 2~ab$^{-1}$ at 250~GeV centre-of-mass (CoM) energy. 
For a 550~GeV collider, a separate optimisation shows that the best solution is close to that obtained by equal scaling of the $\rm{e}^-$ and $\rm{e}^+$ arms of the 250~GeV facility. 

\begin{figure}[h]
\begin{minipage}{0.61\textwidth}
    \includegraphics[width=\textwidth]{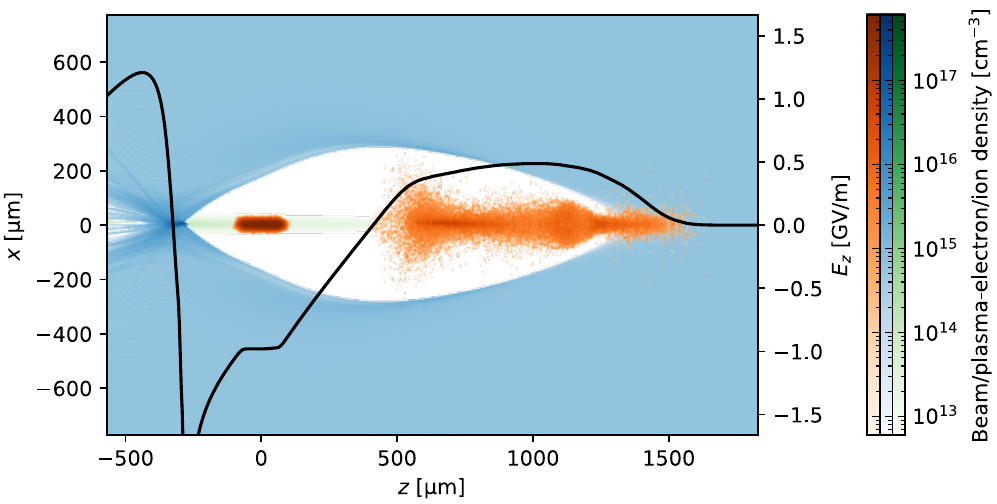}
    \caption{PIC simulation of the accelerating structure proposed for HALHF, with an 8.0/1.6~nC driver/trailing bunch (orange colour map) and a cold helium plasma (blue) of density 6$\times$10\textsuperscript{14}~cm\textsuperscript{--3}. A ramped driver current results in a flattened decelerating field and an 80\% energy-depletion efficiency. Ion motion (green) suppresses the beam-breakup instability.}
    \label{fig:pwfa_sim}    
\end{minipage}
\hfill
\begin{minipage}{0.37\textwidth}
    \centering
    \includegraphics[width=\textwidth]{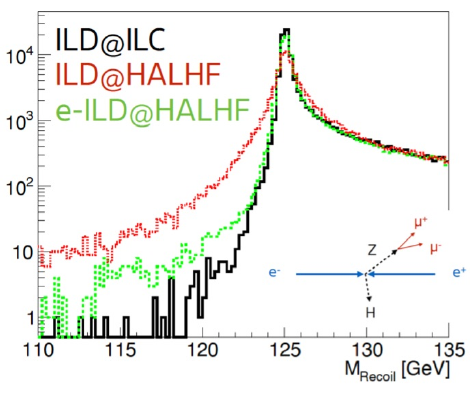}
    \caption{Reconstructed recoil-mass distribution for \ee$\rightarrow$ZH$\rightarrow$\murm{\murm}H events in the ILD detector at HALHF (red, using the original boost of 2.13) compared to that at ILC (black). Elongation of the barrel (green) by 2$\times$ mostly recovers the resolution.}
    \label{fig-4}    
\end{minipage}
\end{figure}

One of the key parameter spaces explored in the Bayesian optimisation is the balance between the power requirements (and thereby the number of klystrons plus modulators, which is the largest cost driver) of the drive-beam linac compared to the linac accelerating the colliding positron bunch. Another is the length, directly proportional to the assumed accelerating gradient, of the linacs. The gradient of the PWFA arm is not a strong cost driver, so a reduction in the plasma density from $7\times10^{15}$~cm$^{-3}$ to $6\times10^{14}$~cm$^{-3}$ has been made (see Fig.~\ref{fig:pwfa_sim}). This greatly reduces performance requirements on many aspects of the PWFA accelerator, which is the most novel element of the facility. This is a very conservative choice; once experience is gained with operating a PWFA accelerator in collider mode, it should be possible to increase the gradient by e.g.\ using a higher plasma density. The optimum found by the Bayesian optimiser for the  baseline has an electron energy of 375~GeV and a positron energy of 41.7~GeV, corresponding to a boost of 1.67 (compared to the original 2.13). 

The optimisation of the drive-beam linac quickly converges to very close to the CLIC drive beam~\cite{CLIC_CDR_2012}. 
This is reassuring, as CLIC has had many person-years of effort \cite{CTF3} devoted to it and the requirements are similar for HALHF. These linacs can be thought of as one arm of a transformer that converts low-energy, high-current electron bunches to high-energy, low-current bunches. The charge in each drive bunch in the HALHF linac is 8~nC at 3.9~GeV. The similarity between the new HALHF baseline and CLIC extends to the addition in HALHF of a delay loop plus combiner rings, with the same functions, \textit{viz.} to produce the required bunch pattern while reducing the peak power load on the linac to an acceptable level. The linac has an average gradient of $\sim$3~MV/m, higher than that of CLIC of $\sim$1~MV/m, and runs with 1~GHz klystrons and modulators. A longer drive-bunch separation and a higher beam energy results in the combiner rings being larger than those of CLIC. The two combiner rings are assumed to be in the same tunnel and interleave in two steps of $3\times$ and $4\times$, giving an overall combination factor of 12. Every fourth RF bucket is filled. Other parameters can be seen in Table~\ref{tab:1}.

\begin{table*}
    \tiny
    \centering
    \vspace{-1cm}
    \begin{tabular}{p{0.24\linewidth}>{\centering}p{0.06\linewidth}>{\centering}p{0.08\linewidth}>{\centering\arraybackslash}p{0.08\linewidth}>{\centering\arraybackslash}p{0.08\linewidth}>{\centering\arraybackslash}p{0.08\linewidth}>{\centering\arraybackslash}>{\centering\arraybackslash}p{0.08\linewidth}>{\centering\arraybackslash}p{0.08\linewidth}}
    
    \textit{Machine parameters} & \textit{Unit} & 
   \multicolumn{2}{c}{\textit{Value (250 GeV)}}& \multicolumn{2}{c}{\textit{Value (380 GeV)}} & \multicolumn{2}{c}{\textit{Value (550 GeV)}}\\
    \hline
    Centre-of-mass energy & GeV & \multicolumn{2}{c}{250} & \multicolumn{2}{c}{380} &\multicolumn{2}{c}{550}\\
    Centre-of-mass boost & ~ & \multicolumn{2}{c}{1.67} & \multicolumn{2}{c}{1.67} & \multicolumn{2}{c}{1.67}\\
    Bunches per train & & \multicolumn{2}{c}{160} & \multicolumn{2}{c}{160}  & \multicolumn{2}{c}{160}  \\
    Train repetition rate & Hz & \multicolumn{2}{c}{100} & \multicolumn{2}{c}{100} & \multicolumn{2}{c}{100}\\
    Average collision rate & kHz & \multicolumn{2}{c}{16} & \multicolumn{2}{c}{16} & \multicolumn{2}{c}{16}\\
    Luminosity & cm\textsuperscript{--2}~s\textsuperscript{--1} & \multicolumn{2}{c}{1.2$\times$10\textsuperscript{34}} & \multicolumn{2}{c}{1.7$\times$10\textsuperscript{34}} & \multicolumn{2}{c}{2.5$\times$10\textsuperscript{34}} \\
    Luminosity fraction in top 1\% & & \multicolumn{2}{c}{63\%} & \multicolumn{2}{c}{53\%} &\multicolumn{2}{c}{46\%} \\
    Quantum parameter ($\Upsilon$) & & \multicolumn{2}{c}{0.9} & \multicolumn{2}{c}{1.6} &\multicolumn{2}{c}{2.8} \\
    Estimated total power usage & MW & \multicolumn{2}{c}{106} & \multicolumn{2}{c}{154} & \multicolumn{2}{c}{218} \\
        Total site length & km & \multicolumn{2}{c}{4.9} & \multicolumn{2}{c}{6.5} & \multicolumn{2}{c}{8.4} \\
    \hline
     & & & \\[-6pt]
    \textit{Colliding-beam parameters} & & $\rm{e}^-$ & $\rm{e}^+$ & $\rm{e}^-$ & $\rm{e}^+$ & $\rm{e}^-$ & $\rm{e}^+$ \\ 
    \hline
    Beam energy & GeV & 375 & 41.7 & 570 & 63.3 & 825 & 91.7 \\
    Bunch population & 10\textsuperscript{10} & 1 & 3 & 1 & 3 & 1 & 3 \\
    Bunch length in linacs (rms) & {\textmu}m & 40 & 150 & 40 & 150 & 40 & 150 \\ 
    Bunch length at IP (rms) & {\textmu}m & \multicolumn{2}{c}{150} & \multicolumn{2}{c}{150} & \multicolumn{2}{c}{150}  \\
    Energy spread (rms) & \% & \multicolumn{2}{c}{0.15} & \multicolumn{2}{c}{0.15} & \multicolumn{2}{c}{0.15} \\
    Horizontal emittance (norm.) & mm mrad & 90 & 10 & 90 & 10 & 90 & 10 \\
    Vertical emittance (norm.) & mm mrad & 0.32 & 0.035 & 0.32 & 0.035 & 0.32 & 0.035\\
    IP horizontal beta function & mm & \multicolumn{2}{c}{3.3} & \multicolumn{2}{c}{3.3} & \multicolumn{2}{c}{3.3}\\
    IP vertical beta function & mm & \multicolumn{2}{c}{0.1} & \multicolumn{2}{c}{0.1} & \multicolumn{2}{c}{0.1} \\
    IP horizontal beam size (rms) & nm & \multicolumn{2}{c}{636} & \multicolumn{2}{c}{519} & \multicolumn{2}{c}{429} \\
    IP vertical beam size (rms) & nm & \multicolumn{2}{c}{6.6} & \multicolumn{2}{c}{5.2} & \multicolumn{2}{c}{4.4} \\
    Average beam power delivered & MW & 9.6 & 3.2 & 14.6 & 4.9 & 21.1 & 7.0 \\
    Bunch separation & ns & \multicolumn{2}{c}{16} &\multicolumn{2}{c}{16} &\multicolumn{2}{c}{16} \\
    Average beam current & {\textmu}A & 26 & 77 & 26 & 77 & 26 & 77  \\[2pt]
    \hline \\[-6pt]
    \multicolumn{2}{l}{\textit{Positron cool-copper RF linac parameters (S-band)}}  & & & & \\
     \hline
     Average cavity gradient & MV/m & \multicolumn{2}{c}{40}& \multicolumn{2}{c}{40} & \multicolumn{2}{c}{40}\\
        Average gradient & MV/m & \multicolumn{2}{c}{36} & \multicolumn{2}{c}{36} & \multicolumn{2}{c}{36}\\
    Wall-plug-to-beam efficiency & \% & \multicolumn{2}{c}{11} & \multicolumn{2}{c}{11} & \multicolumn{2}{c}{11}\\
    RF power & MW & \multicolumn{2}{c}{11.7} &\multicolumn{2}{c}{17.8} &\multicolumn{2}{c}{25.8} \\
    Cooling power  & MW & \multicolumn{2}{c}{17.9} &  \multicolumn{2}{c}{27.3} & \multicolumn{2}{c}{39.5}\\
     Total power & MW & \multicolumn{2}{c}{29.6} & \multicolumn{2}{c}{45.1} & \multicolumn{2}{c}{65.3} \\
   Klystron peak power & MW & \multicolumn{2}{c}{67} & \multicolumn{2}{c}{67} & \multicolumn{2}{c}{67} \\
     Number of klystrons & ~ & \multicolumn{2}{c}{321} & \multicolumn{2}{c}{452} & \multicolumn{2}{c}{678} \\
    RF frequency & GHz & \multicolumn{2}{c}{3} & \multicolumn{2}{c}{3} & \multicolumn{2}{c}{3} \\
 Operating Temperature & K & \multicolumn{2}{c}{77} & \multicolumn{2}{c}{77} & \multicolumn{2}{c}{77} \\
         Length (after damping ring, starting at 3 GeV) & km & \multicolumn{2}{c}{1.1} & \multicolumn{2}{c}{1.7} & \multicolumn{2}{c}{2.5} \\
    \hline \\[-6pt]
    \multicolumn{2}{l}{\textit{Driver linac RF parameters (L-band)}} & & & & \\
    \hline \\[-6pt]
        Average cavity gradient & MV/m & \multicolumn{2}{c}{4} & \multicolumn{2}{c}{4} & \multicolumn{2}{c}{4}\\
    Average gradient & MV/m & \multicolumn{2}{c}{3} & \multicolumn{2}{c}{3} & \multicolumn{2}{c}{3} \\
    Wall-plug-to-beam efficiency & \% & \multicolumn{2}{c}{55} & \multicolumn{2}{c}{55} & \multicolumn{2}{c}{55} \\
    RF power usage & MW & \multicolumn{2}{c}{42.9} & \multicolumn{2}{c}{66.0} & \multicolumn{2}{c}{96.4}\\
   Klystron peak power & MW & \multicolumn{2}{c}{21} &  \multicolumn{2}{c}{21} & \multicolumn{2}{c}{21}\\
     Number of klystrons & ~ & \multicolumn{2}{c}{409} & \multicolumn{2}{c}{630} & \multicolumn{2}{c}{919} \\
    RF frequency & GHz & \multicolumn{2}{c}{1} & \multicolumn{2}{c}{1} & \multicolumn{2}{c}{1} \\
        Length & km & \multicolumn{2}{c}{1.3} & \multicolumn{2}{c}{1.9} & \multicolumn{2}{c}{2.8} \\
\hline \\[-6pt]
\multicolumn{2}{l}{\textit{Combiner ring parameters}} 
    & & \\
    \hline \\[-6pt]
    ~ & ~ & ~ & ~ & ~ & ~ \\[-8pt]
    Delay loop length & m & \multicolumn{2}{c}{1.5} &\multicolumn{2}{c}{1.5} &\multicolumn{2}{c}{1.5} \\
    CR1 diameter & m & \multicolumn{2}{c}{244} &\multicolumn{2}{c}{244} &\multicolumn{2}{c}{244} \\
    CR2 diameter & m & \multicolumn{2}{c}{244} &\multicolumn{2}{c}{244} &\multicolumn{2}{c}{244} \\
    \hline \\[-6pt]
\multicolumn{2}{l}{\textit{PWFA linac and drive-beam parameters}} 
    & & & & \\
    \hline \\[-6pt]
    Number of stages &  & \multicolumn{2}{c}{48} & \multicolumn{2}{c}{48} & \multicolumn{2}{c}{48} \\
    Plasma density & cm\textsuperscript{--3} & \multicolumn{2}{c}{6$\times$10\textsuperscript{14}} & \multicolumn{2}{c}{6$\times$10\textsuperscript{14}} & \multicolumn{2}{c}{6$\times$10\textsuperscript{14}}\\ 
    In-plasma accel. gradient & GV/m & \multicolumn{2}{c}{1}& \multicolumn{2}{c}{1} & \multicolumn{2}{c}{1} \\
    Av. gradient (incl. optics) & GV/m & \multicolumn{2}{c}{0.33}& \multicolumn{2}{c}{0.38} & \multicolumn{2}{c}{0.43} \\
    Transformer ratio & ~ & \multicolumn{2}{c}{2} & \multicolumn{2}{c}{2} & \multicolumn{2}{c}{2} \\
    Length per stage & m & \multicolumn{2}{c}{7.8} &\multicolumn{2}{c}{11.8} & \multicolumn{2}{c}{17.1} \\
    Energy gain per stage\tnote{a} & GeV & \multicolumn{2}{c}{7.8} & \multicolumn{2}{c}{11.8} & \multicolumn{2}{c}{17.1} \\
    Initial injection energy & GeV & \multicolumn{2}{c}{3} & \multicolumn{2}{c}{3} & \multicolumn{2}{c}{3}\\
    Driver energy & GeV & \multicolumn{2}{c}{4}  & \multicolumn{2}{c}{5.9} & \multicolumn{2}{c}{8.6}\\
    Driver bunch population & 10\textsuperscript{10} & \multicolumn{2}{c}{5.0} & \multicolumn{2}{c}{5.0} & \multicolumn{2}{c}{5.0} \\
    Driver bunch length (rms)  & {\textmu}m &  \multicolumn{2}{c}{253} &  \multicolumn{2}{c}{253} & \multicolumn{2}{c}{253} \\
    Driver average beam power & MW & \multicolumn{2}{c}{23.8} & \multicolumn{2}{c}{36.2} & \multicolumn{2}{c}{52.6} \\
    Driver bunch separation & ns & \multicolumn{2}{c}{4} & \multicolumn{2}{c}{4} & \multicolumn{2}{c}{4} \\
    Driver-to-wake efficiency & \% & \multicolumn{2}{c}{80} & \multicolumn{2}{c}{80} & \multicolumn{2}{c}{80} \\
    Wake-to-beam efficiency & \% & \multicolumn{2}{c}{50} & \multicolumn{2}{c}{50} & \multicolumn{2}{c}{50} \\
    Driver-to-beam efficiency & \% & \multicolumn{2}{c}{40} & \multicolumn{2}{c}{40} & \multicolumn{2}{c}{40} \\
    Wallplug-to-beam efficiency & \% & \multicolumn{2}{c}{22} & \multicolumn{2}{c}{22} & \multicolumn{2}{c}{22} \\
    Cooling req.~per stage length & kW/m & \multicolumn{2}{c}{38.4} & \multicolumn{2}{c}{38.4} & \multicolumn{2}{c}{38.4} \\
    Length & km & \multicolumn{2}{c}{1.1} & \multicolumn{2}{c}{1.5} & \multicolumn{2}{c}{1.9} \\
    \hline
    \end{tabular}
   \caption{HALHF parameters for the updated baseline designs at 250~GeV, 380~GeV and 550~GeV CoM energies.}
   \label{tab:1}
\end{table*}

The PWFA linac has a larger number of lower-energy stages than the original baseline: 48 compared to 16. Such a longer train at higher average current but lower final driver energies is more cost effective overall.
The gradient in the plasma cells is conservatively chosen to be 1~GV/m, and the transformer ratio to be 2, which means each stage is 7.8~m long. The drive beams are distributed to each cell in synchronisation with the accelerated bunch by a system of RF deflectors in which drive bunches emerging from the combiner rings are alternately distributed to each side of the array of plasma cells.

The required timing precision to ensure correct drive-beam injection relative to the beam to be accelerated is $\sim$10~fs. Such precision is achievable with state-of-the-art synchronisation; however, the details of this scheme remain to be finalised.
The spent drive beams, which have a near-100\% energy spread but heavily peaked at low energies (a few 100~MeV), are extracted to beam dumps after each cell. 

Since the positron charge is high ($\sim$4.8~nC) and we want to maximise the RF gradient, the positron linac is ideally S-band. In principle, any suitable RF technology could be used; a good option that provides high gradient at only moderate risk is copper structures cooled to liquid-nitrogen temperatures, as developed by the C\textsuperscript{3} Collaboration~\cite{C3}, operated at low gradient (40~MV/m) when compared to the final goal of C\textsuperscript{3} (>150~MV/m). This choice reduces the length of the linac, which otherwise would be the longest single item in the HALHF facility. The choice of cooled copper adds complication in that liquid-nitrogen cryogenics is required, with a power of 16.6~MW (at 250 GeV CoM energy), with concomitant cryogenic plants on the surface and the safety implications of liquid nitrogen in the tunnel. 
It provides a simple path to increasing the positron energy for higher CoM running (see Sec.~\ref{sec:upgrades}). A fallback option would use a warm linac with a gradient of 25~MV/m, similar to the SLAC linac. This would add approximately 1~km to the 250~GeV baseline. 

The beam-delivery systems (BDS) are scaled from the ILC 500 GeV design. Various novel schemes for the BDS are under consideration, which will hopefully either reduce its length or allow it to cope with higher-energy upgrades for the same length. This is particularly important for the electron arm, where collimation between plasma cells could well play an important and beneficial role. Twin beam-delivery systems are envisaged, allowing a sharing of luminosity between two detectors but adding to the overall length. Both BDS will contain spin rotators to produce longitudinal polarization from the transverse polarization that must be maintained in both positron and electron linacs. It is intended that the BDS systems will service two full-scale detectors, with the likely required number of experimenters per detector as discussed in the LCVision submission~\cite{LCVision}. Currently it is envisaged that the detectors would share a single hall but a final decision will be taken when the detailed BDS design, currently under way, is completed. The design of the BDS is more complex than that for CLIC because of the relatively high emittance allowed for the PWFA arm~\cite{Cilento2021}.

As remarked earlier, some elements of HALHF have hardly been studied due to lack of resources. These include the positron damping rings, where it has been assumed that variants on the CLIC design can be used, and the colliding-electron and driver sources. For the electrons, which for HALHF have relaxed emittance requirements compared to ILC and CLIC, it has been assumed that R\&D currently underway for other projects can succeed in developing e.g.~robust and long-lived polarized electron sources that can give the required HALHF beam parameters without the necessity for a damping ring~\cite{Maxon2024}. 
Should this  not  be the case, a damping ring can be added at a relatively minor cost. 

Initial placement studies have been conducted at CERN to find a suitable location for HALHF; this has informed the project cost. The Injection Complex (Driver Linac, Delay Loop, Combiner Ring) could be housed on existing land on the CERN Pr\'{e}vessin campus, in cut-and-cover tunnels. A transfer tunnel connects this surface facility to the accelerator tunnel about 125~m below the surface. This tunnel alignment would be very similar to that for CLIC, all situated in good Molasse Rock. The tunnel diameters have been chosen to be large and therefore conservative; more detailed design work will likely shrink some locations by a substantial fraction. A schematic layout for the civil engineering is shown in Fig.~\ref{Fig-civilstudy}. 

The capital cost and running cost estimates are output by the Bayesian optimisation. The combination of an optimisation of the \textit{total} cost of the project to produce an integrated luminosity of 2~ab\textsuperscript{--1}, and the upgrades outlined above 
lead to the capital cost discussed in Sec.~\ref{sec:costing}.

\subsection{Physics Studies}
\label{sec:physics}
The physics reach of HALHF is similar to that outlined in the LCVision submission~\cite{LCVision}, with differences caused by the boost in the final state inherent in the HALHF concept. GUINEA-PIG~\cite{Schulte_Thesis_1996} simulations have been performed; the calculated luminosities are shown in Table~\ref{tab:1}. At HALHF, all collision events will be boosted into the forward direction, i.e. the direction of the electron beam. Thus, placing a typical symmetric Higgs-factory detector at the HALHF interaction point would lead to significantly worse precision on flagship measurements. Figure~\ref{fig-4} shows a clear and unacceptable deterioration of the recoil mass distribution in \ee$\rightarrow$ZH$\rightarrow$\murm{\murm}H events in the ILD detector placed at HALHF compared to the same detector at ILC. The loss in performance is primarily due to a shortened lever-arm in the solenoidal magnetic field of the detector, and can be recovered e.g. by lengthening the barrel part of the detector by a factor two. This is not intended to be a serious proposal for HALHF, merely to indicate that a properly designed custom detector can recover similar resolution to that expected at a symmetric Higgs factory.

The boost at HALHF should facilitate vertex reconstruction since the flight distances in the lab system become longer for the same proper lifetime; this is counteracted by a loss of acceptance by more particles being lost down the forward beam-pipe. Quantifying these effects requires an optimised geometry of the beam pipe, the vertex detector and the forward tracking. This is subject to one further constraint, namely the distribution of \ee\ pairs created in the beam--beam interaction. The amount and distribution of the pair background has been studied using GUINEA-PIG . The size of the necessary instrumentation-free region is driven by the charge asymmetry, not the energy asymmetry, of the beams. Furthermore, the bunch lengths and the strength of the detector solenoid, which acts to sweep away the \ee\ pairs, also influence the region that can be instrumented. The  inner and forward parts of a purpose-designed detector for the new baseline of HALHF are currently being implemented in  GEANT4. 

The luminosity measurement at \ee\ colliders classically relies on low-angle Bhabha scattering, i.e.\ back-to-back \ee\ pairs at small polar angles, detected in special high-resolution luminosity calorimeters. 
At HALHF, the Bhabhas will be boosted; a detailed simulation study is being conducted to verify the performance and evaluate requirements on the luminosity detection.

Reduction in the CoM energy of HALHF to run at the Z\textsuperscript{0} can be achieved by either reducing the energy in the electron arm to $\sim$46 GeV  or by equally scaling down the energy of both arms to $\sim$15 and $\sim$138~GeV  for the positron and electron arms, respectively. Luminosity is small compared to a circular machine but such running is useful for calibration and polarized beams allow control of systematics. 

\section{Readiness \& necessary R\&D programme}
\label{sec:RandD}

Unlike ILC, HALHF is not ``shovel-ready"; it requires an R\&D programme of $\sim$15~years. It  can be conveniently divided into three parts: RF facilities; those related to the PWFA arm; and the detector. 
Much progress has already happened in the former; most  subsystems are already well-established as they are identical or very close to those used in CLIC, ILC, or under development for C\textsuperscript{3}. For the PWFA arm, although it is much less well developed,  
great strides have been made over the past 5--10 years, particularly in beam  efficiency and quality as well as in user applications and reliability~\cite{LindstromThevenet_2022, Lindstrom2021, Maier2020, Lindstrom2024, Pena2024}. 

In addition to the above ``core'' R\&D programme, other R\&D could improve HALHF's physics reach and/or reduce the cost, e.g.\ structure 
wakefield acceleration (SWFA)\cite{CJing2016, Foster2025}, which is equally capable of accelerating electrons and positrons. Using SWFA to avoid the expense and upkeep of klystrons and modulators by exploiting an already existing beam to power the positron accelerator is attractive. Options include a two-structure beam transformer similar to CLIC as well as a collinear SWFA accelerator~\cite{Zholents2018,Jing2022}.

\subsection{Non-PWFA accelerator R\&D}
\label{sec:CRandD}
The design of the polarized positron source is critical. Although the ILC design, on which the HALHF source is based, is well advanced, a number of technical questions remain, in particular for the rotating target. In addition, the adaption of the undulator to the larger electron energy and the improvement of the capture efficiency to obtain the requisite number of positrons (more than 3 $\rm{e}^+$ per $\rm{e}^-$) require a dedicated effort. The electron source also requires R\&D to develop a robust polarized low-emittance source that can give the required properties without a damping ring.
Since this has been underway for many years for other applications, it is assumed that HALHF can largely piggy-back on these developments.

Designs for both the drive-beam linac, based on CLIC, and the positron linac, based on C\textsuperscript{3} technology, must be developed over the next few years. We have established good relationships with both teams. Optimisation of the beam-delivery system, particularly for higher energies, will be required.

\begin{table}
\scriptsize
\vspace{-1cm}
    \begin{tabular}{p{0.6\linewidth}>
    {\centering}p{0.035\linewidth}>
    {\centering}p{0.05\linewidth}>
    {\centering}p{0.09\linewidth}>
    {\centering}p{0.085\linewidth}}%
    \hline
    {\bf R\&D element}&Start\\year&Duration\\(years) &Personnel\\(FTE years)& Capital\\(MCHF) \cr
    \hline
    \textit{Phase 1: Basic R\&D and integrated collider design} &  0 & 5 & &  \cr
    \hline
    Plasma accelerator R\&D:&  &  & &  \cr
    \tabindent Single-stage quality preservation at high efficiency demonstration (ongoing) & 0 & 5 & 90 & 13 \cr
    \tabindent Basic staging and beam-quality R\&D:&  &  & &  \cr
    \tabindent \tabindent Self-consistent PWFA staging simulations (incl. spin polarization) & 0 & 3 & 15 & 2 \cr
    \tabindent \tabindent Demonstrating achromatic staging optics (nonlinear plasma lens) & 0 & 3 & 5 & 3 \cr
    \tabindent \tabindent Quality-preserving stage-to-stage transport experiment & 3 & 2 &10 & 5\cr
    \tabindent Basic plasma heating and cooling R\&D&  &  & &  \cr
    \tabindent \tabindent Self-consistent long-term plasma evolution simulations & 0 & 3 & 10 & 1\cr
    \tabindent \tabindent Cooled plasma-cell development & 0 & 5 & 30 & 5 \cr
    \tabindent \tabindent High-peak-power plasma evolution experiment (in existing PWFA facilities) & 0 & 5 & 10 & 2 \cr
    Collider design (toward CDR): &  &  & &  \cr
    \tabindent Polarized $\rm{e}^+$ source R\&D & 0 & 5 & 16 & 3\cr
    \tabindent Positron linac design (e.g., cool copper) & 0 & 5 & 10 & 10 \cr
    \tabindent Drive-beam complex design (linac, combiner rings, etc.) & 0 & 5 & 20 & 7\cr
    \tabindent Beam-delivery system design (incl. double IP) & 0 & 5 & 10 & 1 \cr
    \tabindent Asymmetric physics and detector design (Not included in accelerator design) &  &  & & \cr
    \hline
    \textit{Phase 2: Key demonstrations} & 5 & 5 & &  \cr
    \hline
    Plasma demonstrations: & &  & &  \cr
    \tabindent Staging and stability demonstrator (new dedicated facility; SFQED application) & 5 & 5 & 25 & 60 \cr
    \tabindent High-average-power plasma-cell demonstration (upgrading an existing facility) & 5 & 5 & 20 & 20 \cr
    \tabindent Single-stage polarization preservation experiment & 7 & 3 & 15 & 20 \cr
    \tabindent Self-consistent full-train start-to-end simulations & 5 & 5 & 5 & 1 \cr
    Other systems demonstrations: & &  & &  \cr
    \tabindent Cool-copper RF linac demonstrator (parallel development to HALHF) &  &  & &  \cr
    \tabindent Polarized positron source demonstrator (parallel development to HALHF) &  &  & &  \cr
    \hline
    \textit{Phase 3: HALHF (all in one) demonstrator} & 10 & 5 & & \cr
    \hline
    Upgrade of staging facility with high-power plasma cells and RF & 10 & 5 & 20  & 25 \cr
    Upgrade to include increased beam quality & 12 & 3 & 20 & 25  \cr
    Upgrade to include spin polarized source & 13 & 2 & 10 & 10 \cr
    \hline
    \textbf{Total} &  & 15 yrs & 341 FTE yrs & 213 MCHF \cr
    \hline
    \end{tabular}
   \caption{HALHF R\&D Milestones and required resources to produce a Technical Design Report (TDR).}
      \label{tab:2}
\end{table}

\subsection{PWFA R\&D}
\label{sec:PWFA}

The two most critical PWFA research topics are: (1) the demonstration of stable staging with high efficiency and simultaneous emittance preservation and low energy spread; and (2) understanding the limits of repetition rate defined by plasma heating, and how to operate high-average-power plasma cells. 
However, there is also still much work to do to understand and control acceleration within a single plasma stage, particularly simultaneous preservation of charge, emittance, and energy spread. The dependence of beam quality on the bunch pattern and rate must also be investigated. There are several well-known transverse plasma instabilities that must be understood and controlled \cite{Finnerud2025}. Preservation of electron polarization during the plasma acceleration process must also be investigated, as must preservation of the "flat beam" aspect ratio during acceleration \cite{Diederichs2024}. These latter three effects are thought to be understood and can be maintained satisfactorily in simulations but must be confirmed experimentally. 

For staging, recent advances have provided a concept for achromatic transport between stages, based on non-linear plasma lenses~\cite{Drobniak2025}. Such lenses are being developed and tested at the CLEAR facility at CERN. The first full-lattice demonstrations will likely be performed at low energy (sub-GeV), minimising space requirements in the beam line, where both LWFA and PWFA can make valuable contributions. Finally, simulation work toward designing a driver-distribution system, which distributes drivers from the drive-beam linac to each of the plasma cells, is ongoing. 

The operation of the plasma source at high repetition rate and high average power has many challenges. Firstly, it is expected that the plasma itself will be significantly altered between colliding bunches by the energy deposition from all preceding bunches in the train. This cumulative energy deposition, which is orders of magnitude higher than the state of the art, will lead to large thermal velocities of the plasma electrons and ions, modifying the properties of the plasma and thus the acceleration. Research into the nature of energy deposition/evolution in the plasma and its temperature-based effects on the acceleration process is required. Furthermore, a significant proportion of the heat generated by the HALHF bunch pattern will ultimately make its way to the solid materials housing the plasma. This heat must be removed efficiently, likely requiring novel temperature-stabilised designs. 

Underpinning these topics is the development of simulation tools, both the refinement of particle-in-cell (PIC) codes and simplified models to allow collider simulations at reasonable computational cost. The PIC codes, originally designed to describe the first plasma periods in full detail, must be expanded to include additional physics processes relevant to the long-term evolution of the plasma (e.g.~radiative cooling and collisional effects). Moreover, integrated collider simulations require a start-to-end framework; a new code (ABEL) is being developed.  Although some of this work is under way, additional resources are required to achieve the necessary goals.

\subsection{Detector}
\label{sec:DRandD}

 Much of the necessary detector R\&D for HALHF is already underway or has been completed for other designs such as ILC, CLIC, and FCC-ee. However, some aspects are unique to HALHF. In particular, GEANT4 and the fast Monte Carlo suite have to be able to accommodate an asymmetric detector. These must be used to explore design questions such as the appropriate magnetic field configuration to optimise resolution in the forward (boosted) direction, together with designing and if necessary developing detectors that optimise resolution in the forward direction. Luminosity monitors to deal with the different situation in HALHF must be designed and developed. The machine--detector interface is of primary importance as there will certainly be differences compared to a symmetrical machine.

 \subsection{Timeline and Milestones}
 \label{sec:TandM}

Tables~\ref{tab:2} and \ref{tab:3} show estimates of R\&D timelines and resources required to produce a HALHF TDR and demonstrator; Table~\ref{tab:3} includes Technology Readiness Levels (TRL). Some tasks, particularly those more applicable to plasma acceleration in general, e.g.\ the development of a single quality-preserving and energy-efficient plasma cell and staging, already have substantial resources devoted to them (e.g. FACET-II~\cite{FACETII}, FLASHForward~\cite{Darcy2019}, AWAKE~\cite{AWAKE} and ~EuPRAXIA~\cite{Assmann2020}). Those specific to HALHF tabulated here, however, require dedicated funding. For a new technology such as PWFA, it will be essential to build a demonstrator to investigate reliability, operational performance over the medium term and other aspects of a user-based facility. The EU-funded SPARTA project is dedicated to designing a facility for a PWFA-driven $\sim$50~GeV electron beam that can be used for strong-field quantum electrodynamics (SFQED) experiments~\cite{SPARTA}. The concepts developed in the SPARTA project (e.g.~\cite{Drobniak2025}) would be applicable to a demonstrator for HALHF. 

A three-phase R\&D plan is laid out in Table~\ref{tab:2}, aiming to deliver a complete HALHF TDR and demonstrator. Phase~1 concentrates on improving the basic building blocks of a collider to meet HALHF requirements, lasting for $\sim$5 years. It requires the bulk of the resources, 170~FTE-years for plasma-based research, 56~FTE-years for the remainder and $\sim$50~MCHF of capital investment. Successful completion leads to Phase~2, in which specialised demonstrators, including a medium-energy (e.g.~50~GeV) staging demonstrator, show the necessary performance to allow the construction of the Phase-3 demonstrator, upgrading the Phase-2 staging demonstrator. In this, all key aspects of HALHF can be tested simultaneously. Phase~2 would last for $\sim$5~years and require 100~MCHF of capital spend and 65~FTE-years, while the final demonstrator would take $\sim$5~years and require 60~MCHF and 50~FTE-years. 

The most challenging aspects of the R\&D include: understanding plasma-heating effects; designing a cooled plasma cell; suppressing transverse instabilities; staging between plasma cells; generation of sufficient polarized positrons; and reducing jitter sufficiently to produce the required luminosity. 

\begin{table}[h]
\tiny
    \begin{tabular}{>{\raggedright}p{0.17\linewidth}>
    {\centering}p{0.02\linewidth}>
    {\raggedright}p{0.08\linewidth}>
    {\centering}p{0.08\linewidth}>
    {\centering}p{0.10\linewidth}>
    {\centering}p{0.04\linewidth}>
    {\centering}p{0.05\linewidth}>
    {\raggedright}p{0.26\linewidth}}%
    \hline
Critical parameters                                               & TRL    & R\&D time (y) (design/total)                                                      & R\&D current (M€) & R\&D needed \\(M€; design/total)& FTE current & FTE-yrs needed & Comments                                 \cr
    \hline
Electron beams \textgreater{}~100~GeV & 1   & 7--8 / 11--13 &                                                                       & 7/100                                                & 1  & 40  & No PWFA-test facilities have produced \textgreater 100 GeV beams                                                      \cr
    \hline
Acceleration in one stage ($\sim$10~GeV)   & 5    & 5 / 9--10 &                                                                       & 10/100                                              & 3                                & 50                                                                   & AWAKE demonstration but  technology may not be suitable                                             \cr
    \hline
Plasma uniformity \\(long \& trans.)                                & 4      & 5/9--10    &                                                                       & 2/100                                                & 2                                                                       & 15                                                                   & AWAKE demonstration but  technology may not be suitable                                                              \cr
    \hline
Preserving beam quality/emittance                      & 3.5      & 7--8/11--13 & 0.5 (ERC + Oslo national)                                & 3 / 100                                              & 5                      & 25                                                                   & Normalized emittance preserved at \textless 3 um levels with small currents\cr
    \hline
Spin, polarization                                                & 2                                                                       & 5 / 9--10    &      0.1 (DESY)                                                                 & 3/100                                                & 1                                                 & 16                                                                   & Technology concept formulated                  \cr
    \hline
Stabilisation (active and passive)                                & 3                                                                        & 7--8 / 11--13 &                                                                       & 1/100                                               & 1                                                                       & 10                                                                   & Studies at AWAKE and LWFA, but not at HALHF requirements                                                          \cr
    \hline
Ultra-low-emittance beams                                         & 2                                                                           & 7--8 / 11--13&                                                                       & 3/100                                                & 0                                                                       & 20                                                                   & Not yet collider emittances; need better test facilities.                                                                                                                   \cr
    \hline
External injection and timing                                     & 4                                                                       & 7--8 / 11--13 &                                                                       & 1/100                                               & 0                                                                       & 10                                                                   & Precise timing for external injection demonstrated at AWAKE                                                                         \cr
    \hline
High rep-rate targetry with heat management                       & 2                                                                      & 5 / 9--10   &                                                                      & 7/100                                                & 3                                               & 40                                                                   & Heat modification of plasma properties/profile and  target cooling requires new concepts                 \cr
    \hline
Temporal plasma uniformity/stability                           & 4                                                                    & 5 / 9--10    &                                                                       & 3/100                                               & 0                                                                       & 10                                                                   & AWAKE demonstration but technology may not be suitable                        \cr
    \hline
Driver removal                                          & 2                                                                         & 7--8 / 11--13 &                                                                       & 2/100                                               & 0.5                                                                     & 10                                                                   & HALHF concept exists                                                     \cr
    \hline
Drivers @ high rep.~rate \& eff.& 5                                                                      & 5 / 9--10   &                                                                       & 5/100                                                & 0.5                                                                     & 10                                                                   & Similar to CLIC driver, demonstrated in CTF3         \cr
    \hline
Interstage coupling                              & 2                                                                & 7--8 / 11--13 & 1 (ERC)                                                        & 3/100                                                & 3                                                                       & 10                                                                   & HALHF concept exists              \cr
    \hline
Total system design with end-to-end                               & 3                                                                      & 3--4                                                       & ~0.5 (Oslo, Oxford)                                           & 3                                                     & 2                                                                 & 20                                                                   & Not yet at pre-CDR level. Aim for pre-CDR document early in 2026.    \cr
    \hline
Simulations                                                       & 5                                                                   & part of above                                                   & 0.5 (ERC + Oslo national)                                & part of above                                                                 & 4                                                 & 5                                                                    & Single-stage simulation well developed - dedicated framework (ABEL) for start-to-end                \cr
    \hline
Self-consistent design                                            & 4                                                                    & part of above                                                   & part of above                                                         & part of above                                                                 & in prev. 2 rows                                            & 5                                                                    & Plasma linac start-to-end simulations performed using HIPACE++/ABEL \cr          
     \hline
\end{tabular}
\caption{HALHF plasma-arm R\&D: Technology Readiness Levels (TRL), required resources and timescales to produce TDR. ``FTE current'' means currently in place; ``needed'' is integrated total requirement.}
      \label{tab:3}
\end{table}

\section{Upgrades}
\label{sec:upgrades}

  HALHF would begin at 250 GeV and be raised in energy by increasing the gradient of the positron linac, the density of plasma and/or the number of plasma cells, with concomitant changes to the number and energy of the drive beams and possibly increased tunnel lengths (see Fig.~\ref{Fig-civilstudy}). HALHF Technology can also be used to upgrade the energy of an already existing CLIC or ILC, as outlined in LCVision~\cite{LCVision}. 

 HALHF offers a path to multi-TeV. The current design could reach  $\sim$1~TeV. However, achieving 3 TeV (CLIC's maximum energy) while maintaining a boost of 1.67 needs 5 TeV electrons and 450 GeV positrons. Additional R\&D, including on emittance preservation, is essential. Recently P5~\cite{P52024} identified 10~TeV parton-CoM collisions as the next step beyond an \ee\ Higgs factory, as addressed by a recent 10-TeV collider effort \cite{10TeVESPP}. If $\rm{e}^+$ acceleration in plasma remains intractable, a 10-TeV $\gamma\gamma$~\cite{Barklow2022} or $\rm{e}^-\rm{e}^-$~\cite{Yakimenko2019} collider should be considered, since the distinction between \ee, $\gamma \gamma$ and $\rm{e}^-\rm{e}^-$ 
 is blurred due to the dominance of vector boson fusion~\cite{Ali2021}. 
 A 10 TeV $\gamma\gamma$ collider is a natural upgrade for HALHF~\cite{HALHF_upgrades}.

R\&D challenges for a 10~TeV linear collider include
beamstrahlung (important even for $\gamma\gamma$), the design of Compton/$\gamma\gamma$ IPs, betatron radiation, jitter tolerances, and BDS at 5~TeV . A test facility with beams of energy >100~GeV will be necessary to investigate these issues. 
The second IP would permit development of the Compton/$\gamma\gamma$ IPs. Lastly, a unique opportunity for a 550~GeV HALHF is to interact the 825~GeV electron beam with a PW laser to reach the fully non-perturbative regime of SFQED~\cite{Fedotov2017}.

\section{Costing}
\label{sec:costing}
The costing for three different CoM energies is detailed in Table~\ref{tab:4}. The similarity between many aspects of HALHF and CLIC allows some costings to be derived therefrom with appropriate scaling~\cite{CLIC_PIP_2018}. This is particularly true of the civil engineering. Other costings, e.g.\ for the klystrons and modulators, are derived from quotations or literature as detailed in the associated Backup document. The basis of estimation is 2012 ILC units, which are converted into 2024~CHF (see Backup). A separate estimate of FTEs for construction has not been attempted; given the scale of the projects, a significantly smaller number than required for either ILC or CLIC is predicted. The costing is understood to $\sim$30\%. Note that underground tunnel costs are conservatively based on 8~m diameter tunnels everywhere ($\sim$2$\times$ volume compared to CLIC) in order to fit the PWFA driver-distribution and cool-copper RF systems; however, an optimisation would significantly reduce the tunnel costs.

\begin{table}[!h]
\scriptsize
    \begin{tabular}{p{0.18\linewidth}>
    {}p{0.35\linewidth}>
    {\centering}p{0.11\linewidth}>
    {\centering}p{0.11\linewidth}>{\centering}p{0.11\linewidth}}
    \hline
    \multirow{2}{8em}{\textit{Domain}} & \multirow{2}{8em}{\textit{Sub-domain}} & \multicolumn{3}{c}{\textit{Cost [MILCU]}} \cr  
   & & \textit{250 GeV} & \textit{380 GeV} & \textit{550 GeV} \cr  
    \hline
    \multirow{6}{8em}{Main-beam production} & Electron source (photocathode, polarized) & 82 & 82 & 82 \cr
        & Electron injector linac & 22 & 22 & 22 \cr
        & Positron source (helical undulator, polarized) & 178 & 178 & 178 \cr
        & Positron injector linac & 32 & 32 & 32 \cr
        & Positron transport & 55 & 74 & 96 \cr
        & Positron damping rings (2x) & 200 & 200 & 200 \cr
    \hline
    \multirow{5}{8em}{Drive-beam production} & Electron source & 10 & 10 & 10 \cr
        & Driver linac modules & 113 & 173 & 254 \cr
        & Driver linac RF & 325 & 501 & 731 \cr
        & Frequency multiplication (combiner rings) & 127 & 127 & 127 \cr
        & Driver transport (surface-to-underground) & 24 & 25 & 26 \cr
    \hline
    \multirow{4}{8em}{Electron linac (PWFA)} & Plasma modules & 17 & 26 & 38 \cr
        & Interstage transport & 30 & 37 & 44 \cr
        & Driver delay chicanes & 90 & 120 & 155 \cr
        & Driver beam dumps & 11 & 17 & 25 \cr
    \hline
    \multirow{3}{8em}{Positron linac\\(cool-copper RF)} & Cool-copper linac modules & 113 & 176 & 259 \cr
        & Cool-copper linac RF & 298 & 465 & 683 \cr
        & LN$_2$ reliquification plants & 34 & 53 & 78 \cr
    \hline
    \multirow{4}{8em}{Beam delivery and post collision lines (dual IPs)} & Electron beam delivery systems (2x) & 158 & 194 & 234 \cr
        & Positron beam delivery systems (2x) & 53 & 65 & 78 \cr
        & Final focus, experimental area & 20 & 20 & 20 \cr
        & Post collision lines/dumps & 45 & 64 & 88 \cr
    \hline
    \multirow{6}{8em}{Civil engineering} & Surface driver and complex & 63 & 92 & 130 \cr
     & Surface-to-underground tunnel & 31 & 31 & 31 \cr
     & Electron arm tunnel & 44 & 59 & 75 \cr
     & Positron arm and damping ring tunnels & 54 & 77 & 106 \cr
     & Beam-delivery systems & 164 & 201 & 243 \cr
     & Interaction region & 154 & 154 & 154 \cr
    \hline
    \multirow{4}{8em}{Infrastructure and services} & Electrical distribution & 104 & 125 & 150 \cr
        & Survey and alignment & 80 & 96 & 116 \cr
        & Cooling and ventilation & 302 & 439 & 622 \cr
        & Transport / installation & 24 & 29 & 35 \cr
    \hline
    \multirow{4}{8em}{Machine control, protection and safety systems} & Safety systems & 30 & 36 & 43 \cr
        & Machine control infrastructure & 60 & 72 & 87 \cr
        & Machine protection & 6 & 7 & 9 \cr
        & Access safety \& control system & 9 & 11 & 14 \cr
    \hline
    \multicolumn{2}{l}{\textbf{Total (in 2012 MILCU)}} & \textbf{3162} & \textbf{4090} & \textbf{5275} \cr
    \hline
    \multicolumn{2}{l}{\textbf{Total} (in 2024 Swiss francs)} & \textbf{3.8~BCHF} & \textbf{4.9~BCHF} & \textbf{6.3~BCHF} \cr 
     \hline \\
    \end{tabular}
   \caption{HALHF costing in MILCUs at 250 GeV, 380 GeV and 550 GeV CoM energies. Overheads ("Infrastructure..." and "Machine...") are scaled from CLIC based on combined civil engineering costs, except "Cooling and ventilation", which is scaled from CLIC based on the total power load.}
      \label{tab:4}
\end{table}

\section{Summary}
The HALHF concept offers for the first time a realistic and timely design for a plasma-based facility that can provide cutting-edge particle physics. Although it requires significant R\&D and the construction of a demonstrator facility over a fifteen-year period, its low carbon and physical footprint, construction cost and long-term perspective to reach the 10~TeV frontier make it an attractive option for investment. The many off-ramps for the necessary R\&D towards applications in other scientific fields of major societal benefit and interest make HALHF an exciting and potentially disruptive development in particle physics.

\newpage
\section*{ADDENDUM}
\label{sec:Addendum}
As requested, we here set out additional information required for large-scale projects. However, given the stage at which HALHF currently finds itself, the information requested can only be partial and in some cases does not exist. We give what information we can under the requested headings.
\renewcommand*{\thesection}{\Alph{section}.}
\renewcommand*{\thesubsection}{\Alph{section}.\arabic{subsection}.}
\renewcommand*{\thesubsubsection}{\Alph{section}.\arabic{subsection}.\arabic{subsubsection}.}
\setcounter{section}{1}
\subsection{Stages and parameters}
\subsubsection{The main stages of the project and the key scientific goals of each}
The scientific goals are identical to those given in the LCVision submission~\cite{LCVision}. Three stages of the project are possible: a Higgs-factory stage at 250~GeV; running at 380 GeV near the top threshold; 550 GeV running for the triple-Higgs coupling etc. Whether these would be constructed in such stages, or whether a single construction phase including a tunnel sufficiently long for 550~GeV would be preferable is a matter for detailed planning at the TDR, EDR and approval stages. 
\subsubsection{Ordering of stages} 
There is some flexibility in the ordering of stages if a 550~GeV tunnel is constructed initially. Otherwise it would proceed without flexibility from lowest to highest energy. 
\subsubsection{Main technical parameters of each stage}
The technical parameters are given in Table~\ref{tab:1} in the main text.
\subsubsection{The number of independent experimental activities and the number of scientists expected to be engaged in each} 
Two interaction points are envisaged with parameters as given in the Backup document. Two general-purpose detector will be constructed, one for each IP. Given the relative complexity of the physics and backgrounds, we would expect the detectors to have a technical complexity between that of the LEP detectors and the LHC detectors, significantly closer to LHC than LEP. Thus the number of experimenters should be close to but smaller than the number engaged on ATLAS and CMS. 
\subsection{Timelines}
\label{sec:timelines}
Before any approval or construction for HALHF can be contemplated, an R\&D phase of some 10-15 years is envisaged. Depending on the success of this R\&D, it will be necessary to construct a demonstrator to ensure that this new technology behaves as anticipated in a user-oriented facility. Construction of such a machine assuming success of the R\&D programme can begin before the 
completion of the collider-oriented parts of the programme. It is therefore feasible to imagine approval and construction of the HALHF facility within about 15 years of the commencement of the fully funded R\&D programme. 
\subsubsection{The technically-limited timeline for construction of each stage}
If HALHF is constructed in a single stage for 550~GeV CoM, the construction time will be only slightly shorter than that envisaged for CLIC, i.e. five years. This is because many of the subcomponents are common, e.g.\ the surface installation of the delay and combiner rings, etc.\ and only the length of the overall tunnel is somewhat shorter for HALHF. For construction in stages starting at 250 GeV, then an initial construction period of around four years is estimated. Further stages to reach to 550 GeV would require further civil construction but some part at least of this could be carried out in parallel with running and/or in long annual shutdown periods. 
\subsubsection{The anticipated operational (running) time at each stage, and the expected operational duty cycle}
The current estimates for the HALHF instantaneous luminosity and operating time are similar to those of the baseline ILC. The physics requirements and operating times at each stage are therefore expected to be similar to those set out in the LCVision document for such a scenario. Given the status of the HALHF design, it would be too ambitious to give estimates for the duty cycle. However, given the fact that the great majority of the HALHF technology is based on linacs and other accelerator elements similar to CLIC, there is no reason to believe that such duty cycles would differ significantly from those of CLIC. 
\subsection{Resource requirements}
\label{sec:resources}
Before approval and construction can begin, a dedicated and funded programme of R\&D is necessary. As detailed in Tables~\ref{tab:2} and \ref{tab:3} above, this amounts in total to approximately 341 FTE years and 213 MCHF over a period of around 15 years. This includes the capital cost of constructing a demonstrator that could be used for e.g.\ strong-field QED experiments, as discussed in the main body of the text. 
\subsubsection{The capital cost of each stage in 2024 CHF}
Details of the capital cost are given in Table~\ref{tab:4} in the main text. In summary, the 250~GeV HALHF facility costs 3.8~BCHF, the 380~GeV version 4.9~BCHF and the 550~GeV facility 6.3~BCHF, in 2024 CHF.  
\subsubsection{The annual cost of operations of each stage}
Since the power consumption for each stage (see Table~\ref{tab:1}) is very similar to that of ILC, and since we have not attempted a detailed breakdown of operations cost, we assume the same costs as those indicated for ILC in the LCVision submission.
\subsubsection{Human resources (in FTE)}
We refer once again to the LCVision submission, although with the proviso that since some fraction of human resources is proportional to the length and therefore the number of components of the facility, and that HALHF is significantly shorter than ILC or CLIC, we would expect the HALHF number of FTEs to be significantly smaller than either ILC or CLIC. 

\subsubsection{Basis-of-estimate of the resource requirements}
The resource requirements have been estimated with the specially developed ``system code"\footnote{A \textit{system code} is a code framework that allows simulation of the physics as well as the engineering layout and cost modelling. An example from fusion reactor research is PROCESS.} ABEL, which produces a broken-down list of elements, costed individually (e.g., klystrons) or by length (e.g., accelerator modules, transfer lines and tunnels). The per-element and per-length costs are based on costs from CLIC and ILC, appropriately adjusted for inflation and exchange rates to be expressed in 2012 dollars (ILC units). Similarly, the power requirements are added up on a per-element basis (largely driven by RF and cooling components), but also adding a non-specific 25\% power overhead for magnets, ventilation etc.~(this will subsequently be estimated in more detail).

\subsection{Environmental impact}
\label{sec:environmental}
\subsubsection{The peak (MW) and integrated (TWh) energy consumption during operation of each stage} 
The estimated peak power is 106~MW for 250~GeV, 154~MW for 380~GeV, and 218~MW for the 550~GeV version of HALHF. Given that this is very similar to that of CLIC and ILC and would follow a similar physics programme,  we refer to the numbers on integrated energy consumption in the LCVision submission.
\subsubsection{The integrated carbon-equivalent energy cost of construction}
At this stage of the project, it would be premature to give a figure for the carbon-equivalent cost of construction, especially as the tunnel diameters have not yet been optimised and have therefore been chosen to be conservatively large. Given the relative footprints of HALHF compared to CLIC, however, we confidently expect the HALHF carbon-equivalent cost to be considerably less than that of CLIC.  
\subsubsection{Any other significant expected environmental impacts}
There are no particularly different environmental impacts of HALHF compared to other linear colliders. The only different elements in HALHF are the cool copper positron linac and the PWFA arm. The cool copper linac will require liquid-nitrogen cooling. The power for such cooling is taken into account in the overall facility power usage and other environmental effects are significantly smaller than the cooling required for ILC. The PWFA requires gas to be ionised but likely candidates include hydrogen, helium, nitrogen, and argon, none of which pose particular environmental challenges in the very small quantities required. A significant amount of heat is generated in the beam dumps as well as short-lived radioactivity. The heat may be sufficient to investigate its usage for heating elsewhere in the campus. The radioactivity poses similar challenges to those at either CLIC or ILC. In general, one of the strengths of HALHF is that overall it has a much smaller environmental impact for construction than any other of the current Higgs-factory proposals. 
\subsection{Technology and delivery}
\label{sec:technology}
\subsubsection{Key technologies}
The key technology for HALHF is beam-driven plasma-wakefield acceleration. Most of the other parts of the facility are relatively standard variations on long-standing accelerator components. The exception is the cool-copper linac for positron acceleration. This is under development in the USA and has yet to be used in a demonstrator. However, this development is proceeding in parallel to HALHF and we are in close contact with the C$^3$ collaboration. The performance parameter required, a gradient of 40~MV/m, is much smaller than the development goal of this technology and therefore relatively safe. However, if it transpires that this technology does not develop as expected, we have a very safe fall-back in terms of a warm SLAC-like linac with a gradient of 25~MV/m. 

The main technology development is that required for PWFA. The key parameters that must be reached in the PWFA arm of HALHF are given in Table~\ref{tab:1}. Although a large community is working on the development of PWFA, the requirements for a PWFA-based collider are well beyond state of the art. Indeed, for some, in particular power and the concomitant cooling requirement, the HALHF requirements are orders of magnitude away from what is currently attainable. To reach these powers, new and innovative plasma-cell designs will be required. Most critical however and a \textit{sine qua non} is the demonstration of the ''staging'' of more than one cell together. HALHF requires no less than 48 cells to be linked, with beam losses and beam-quality degradation reduced to minuscule amounts. As yet, a single demonstration of staging exists but the fraction of the beam transmitted was orders of magnitude away from the HALHF requirements~\cite{Steinke2016}. In addition, beam-quality preservation in the acceleration process must be demonstrated simultaneously in six dimensions rather than only in subsets as is currently the case, the pointing jitter of the beams must be improved by an order of magnitude, etc.
These developments have been carefully considered and a coherent and focused R\&D programme outlined, as shown in Tables~\ref{tab:2} and \ref{tab:3}. 
\subsubsection{The critical path for technology development or design}
The critical path, which depends on some technological developments not directly carried out by the HALHF Collaboration (e.g.\ for the Cool-Copper linac) is as follows:
\begin{enumerate}
    \item Staging demonstration;
    \item Demonstration of single-cell parameters at low power;
    \item Development of a plasma working point suitable for high-average-power operation;
    \item Development of high-average-power plasma cells;
    \item Demonstration of single-cell parameters at high power;
    \item Demonstration of required parameters in two staged cells;
    \item Demonstration of required parameters in mulitple staged cells;
    \item Construction of demonstrator at $\sim$50~GeV and use in SFQED experiments;
    \item Demonstration of sufficient polarised positron production;
    \item Demonstration of polarisation preservation in the PWFA process;
    \item Achievement of 40~MV/m gradients in cool-copper technology demonstrator;
    \item Integrated design of all elements and full-scale simulation leading to a Technical Design Report.
\end{enumerate}
\subsubsection{Key technical risks}
The key technical risks are (not ordered in importance):
\begin{enumerate}
    \item Inability to reach required beam quality and stability in single plasma stages
    \item Inability to achieve staging with required beam loss and quality;
    \item Inability for the plasma properties to be made similar at the arrival time of each bunch in the train;
    \item Inability to design robust plasma cells capable of withstanding the thermal stress of the large power deposition;
    \item Inability to preserve polarisation of the electrons in PWFA arm;
    \item Inability to produce fully integrated and accurate cradle-to-grave simulation of the complete HALHF facility;
\end{enumerate}
There are no mitigating procedures if any of these developments is unsuccessful. Other technical risks for which alternative solutions exist are:
\begin{enumerate}
    \item Inability to achieve 40~MV/m gradient in cooled-copper linac;
    \item Inability to produce the requisite number of polarised positrons.
\end{enumerate}
\subsubsection{Financial and human resources needed for R\&D on key technologies}
These are detailed in Tables~\ref{tab:2} and \ref{tab:3}.
\subsection{Dependencies}
\subsubsection{Whether a specific host site is foreseen, or whether options are available}
Preliminary civil engineering studies have been carried out for a site at CERN along the path of the CLIC site study. It is clear that HALHF could be constructed here. However, HALHF is not limited to a CERN site. Its small footprint means that it could fit within the sites of many current laboratories. If necessary, it could also be constructed at a green-field site. 
\subsubsection{Dependencies on existing or required infrastructure}
There are no dependencies. Assuming that a HALHF demonstrator is constructed somewhere, there would be clear synergies in constructing HALHF at the same site. If a ''conventional'' linear collider were constructed, an upgrade using HALHF technology would clearly benefit from using the same tunnel and could use much of the same infrastructure, particularly if the CLIC technology had already been constructed. 
\subsubsection{The technical effects of project execution on the operations of existing infrastructures at the host site} 
There are no effects except for possible vibrations during construction. 
\subsection{Current project status}
\label{sec:currentstatus}
\subsubsection{Current design / R\&D / simulation activities}
These have been described in Sec.~\ref{sec:CS}. In terms of the community pursuing the project, the current HALHF collaboration is indicated in the footnote to the first page and is listed below. It consists of more than 60 individuals but almost none of them can devote more than a tiny fraction of their time to HALHF. The total FTE count excluding students is probably fewer than five. Other than funding research students, there is essentially no dedicated research funding explicitly for HALHF. However, there are very strong research activities in areas directly applicable to HALHF. These include in general PWFA research, particularly at FACET-II at SLAC, SPARC-LAB at Frascati, AWAKE at CERN and FLASHForward at DESY, and on Cool Copper based mostly in the USA, particularly at SLAC. The EU-funded SPARTA project has very substantial overlap with some elements of HALHF but its deliverables are distinct. 
\subsubsection{Major in-kind deliverables already negotiated}
None. However results from SPARTA (staging), Oxford (plasma heating and cell cooling), and FLASHForward (beam quality, efficiency, high average power) in particular will have direct relevance to HALHF deliverables. 
\subsubsection{Other key technical information points}
None.
\newpage
\noindent
$\dagger$ The HALHF Collaboration: Erik Adli$^1$, Joshua Appleby$^2$, Timothy L. Barklow$^3$, Maria Enrica Biagini$^4$, Jonas Björklund Svensson$^5$, Mikael Berggren$^6$, Simona Bettoni$^7$, Stewart
Boogert$^8$, Philip Burrows$^2$, Allen Caldwell$^9$, Jian Bin Ben Chen$^1$, Vera Cilento$^{2,10}$, Laura Corner$^{11}$, Richard D’Arcy$^2$, Steffen Doebert$^{10}$, Wang Dou$^{12}$, Pierre Drobniak$^1$, Calvin Dyson$^{13}$, Sinead Farrington$^{14}$, John Farmer$^{9}$, Angeles Faus-Golfe$^{15}$, Manuel Formela$^6$, Arianne Formenti$^{16}$, Louis Forrester$^{13}$, Brian Foster$^{2,6}$, Jie Gao$^{12}$, Spencer Gessner$^3$, Niclas Hamann$^6$, Alexander Harrison$^2$, Mark J. Hogan$^3$, Eir Eline H{\o}rlyk$^1$, Maryam Huck$^6$, Daniel Kalvik$^1$, Antoine
Laudrain$^6$, Reme Lehe$^{16}$, Wim Leemans$^6$, Carl A. Lindstr{\o}m$^1$, Benno List$^6$, Jenny List$^6$, Xueying Lu$^{17}$, Edward Mactavish$^{10}$, Vasyl Maslov$^6$, Emilio Nanni$^3$, John Osborne$^{10}$, Jens Osterhoff$^{16}$, Felipe Pe\~na$^{18,1}$,
Gudrid Moortgat Pick$^{19,6}$, Kristjan P\~{o}der$^6$, J\"{u}rgen Reuter$^6$,  Dmitrii Samoilenko$^6$, Nela Sedlackova$^{13}$, Andrei Seryi$^{20}$, Kyrre Sjobak$^1$, Terry Sloan$^{21}$, Rogelio Tomas Garcia$^{10}$, Maxim Titov$^{22}$, Malte Trautwein$^{19}$, Steinar Stapnes$^{10}$, 
Maxence Th\'evenet$^6$, Nicholas J. Walker$^6$, Marc Wenskat$^6$, Matthew Wing$^{23,6}$, Jonathan Wood$^6$.\\
\\
$^1$ Department of Physics, University of Oslo, 0316 Oslo, Norway\\
$^2$ John Adams Institute for Accelerator Science at University of Oxford, Denys Wilkinson Building, Keble Road, Oxford, OX1 3RH, UK\\
$^3$ SLAC National Accelerator Laboratory, 2575 Sand Hill Road, CA 94025, Menlo Park, USA\\
$^4$ Laboratori Nazionali di Frascati, INFN, Via Enrico Fermi, 54, 00044 Frascati RM, Italy\\
$^5$ Department of Physics, Lund University, Box 118, 221 00 Lund, Sweden\\
$^6$ DESY, Notkestrasse 85, 22607, Hamburg, Germany\\
$^7$ Paul Scherrer Institute, Forschungsstrasse 111, 5232 Villigen, Switzerland\\ 
$^8$ Cockcroft Institute, Daresbury Laboratory, STFC, Keckwick Lane, Daresbury, WA4 4AD, Warrington, UK\\
$^{9}$ Max Planck Institut für Physik, Boltzmannstrasse 8, 85748 Garching/Munich, Germany\\
$^{10}$ CERN, CH-1211 Geneva 23, Switzerland\\
$^{11}$ Cockcroft Institute at University of Liverpool, School of Engineering, The Quadrangle, Brownlow Hill, Liverpool L69 3GH, UK\\
$^{12}$ Institute of High Energy Physics, Chinese Academy of Sciences, 9 Yuquan Rd, Shi Jing Shan Qu, Bei Jing Shi, 100039, China.\\ 
$^{13}$ John Adams Institute for Accelerator Science at Department of Physics, Imperial College London, Prince Consort Road, South Kensington, London SW7 2BW, UK\\
$^{14}$ Rutherford Appleton Laboratory, STFC, Harwell Campus, OX11 0QX, Didcot, UK\\
$^{15}$ Laboratoire de Physique des 2 Infinis Irène Joliot-Curie, IJCLab, Orsay, Bât. 100 et 200, 15 rue Georges Clémenceau, F-91405 Orsay, France\\
$^{16}$ Lawrence Berkeley National Laboratory, 1 Cyclotron Rd, Berkeley, CA 94720, USA\\
$^{17}$ Argonne National Laboratory, 9700 S Cass Avenue, IL60439, Lemont, USA\\
$^{18}$ Ludwig-Maximilians-Universität München, Geschwister-Scholl-Platz 1, 80539 München, Germany \\
$^{19}$ II. Institute of Theoretical Physics, University of Hamburg, Luruper Chaussee 149, 22761 Hamburg, Germany \\
$^{20}$ Thomas Jefferson National Accelerator Facility, 12000 Jefferson Avenue, VA23606, Newport News, USA\\
$^{21}$ University of Lancaster, Bailrigg, Lancaster LA1 4YW, UK\\
$^{22}$ CEA-IRFU, Bât. 141 CEA - Saclay, 91191 Gif-sur-Yvette, France\\
$^{23}$ Department of Physics \& Astronomy, University College London, Gower St, London WC1E 6BT, UK\\
\bibliographystyle{elsarticle-harv} 

\end{document}